\documentclass[usenatbib,usegraphicx,referee]{mn2e}
\usepackage{epsfig}
\def\araa{ARA\&A}
\def\mnras{MNRAS}
\def\apj{ApJ}
\def\apjs{ApJS}
\def\aj{AJ}
\def\apjl{ApJL}
\def\aap{A\&A}

\def\baas{BAAS}

\newcommand{\vU}{{\vec{ U}}}
\def\V{{\cal V}}

\def\HI{{\rm HI}}
\begin{document}
\title{The  Scale Height of NGC~1058  Measured from its HI Power
  Spectrum}
\author[Prasun Dutta, Ayesha Begum, Somnath Bharadwaj and Jayaram
  N. Chengalur]{Prasun Dutta$^{1}$\thanks{Email:
    prasun@cts.iitkgp.ernet.in},  
Ayesha Begum$^{2}$\thanks{Email: begum@astro.wisc.edu}, 
Somnath  Bharadwaj$^{1}$\thanks{Email: somnath@cts.iitkgp.ernet.in} and 
Jayaram N. Chengalur$^{3}$\thanks{Email: chengalu@ncra.tifr.res.in}
\\$^{1}$ Department of Physics and Meteorology \&
 Centre for Theoretical Studies, IIT Kharagpur, Pin: 721 302, India, 
\\$^{2}$Department of Astronomy,
University of Wisconsin,
475 N. Charter Street,
Madison, WI 53706 , 
\\$^{3}$ National Centre For Radio Astrophysics, Post Bag 3,
Ganeshkhind, Pune 411 007, India.} 
\maketitle

\begin{abstract}
We have measured the HI power spectrum of the nearly  face-on spiral
galaxy NGC~1058 from radio-interferometric observations using a
visibility based estimator. The power  spectrum is well fitted by
two different power laws $P(U)=AU^{\alpha}$, one  with 
$\alpha\ =-\ 2.5\pm 0.6$ at small length-scales $(600 \, {\rm pc} \,
{\rm to} \,   1.5 \, {\rm kpc})$ and another with  
$\alpha\ =-\ 1.0\pm 0.2$ at large length-scales
$(1.5 \,  {\rm kpc} \, {\rm to} \,  10.0 \, {\rm kpc})$. We interpret
this  change in the slope of the power spectrum as 
 a transition from 3D turbulence at  small
length-scales to 2D turbulence in the plane of the galaxy's disk
at  large length-scales. We use the observed break in the power
spectrum to estimate the galaxy's scale-height,  which 
we find  to be   $  490 \pm 90 $ pc.
\end{abstract}


\section{introduction}
\label{sec:intro}

Evidence has been mounting in recent years that turbulence plays an 
important role in determining the physical conditions of the neutral 
interstellar medium (ISM) as well as generating the hierarchy of 
structures seen in it (see \citealt{ES04I}; \citealt{ES04II}, for
recent reviews). Observational evidence includes the fact that the 
fluctuation power spectrum of a variety of tracers (HI 21-cm emission 
intensity, HI 21-cm optical depth, dust and molecular emission) is a scale-free 
power-law  (\citealt{CD83}; \citealt{GR93}; \citealt{SBH98}; \citealt{SSD99};
  \citealt{DD00};  \citealt{EK01}). 
This scale-free behavior is characteristic of a turbulent medium. 
Similarly, the HI distribution in several dwarf galaxies in the M81 
group appears to be fractal (\citealt{WC99}).  
Further,  the Fourier transform  power spectra of 
 the V and H$\alpha$ images of a sample of
irregular galaxies are  also found to be a power law, indicating that  there is no
characteristic mass or luminosity scale for OB associations and star
complexes (Willett et al. 2005). 

Recently \citet{AJS06} have presented a visibility based formalism for
determining the power spectrum 
of  HI intensity fluctuations in galaxies with
extremely weak emission. This was applied to the dwarf galaxy
DDO~210 and the power spectrum was found to be a power
law with slope $-2.8 \pm 0.4$ over the length scales $100 - 500$
pc. In a subsequent paper, we \citep{DBBC08} have applied  the same
formalism to an external disk galaxy NGC~628 (M~74) and found the 
slope to be $-1.7 \pm 0.2$ over the length scales 
$0.8 - 8.0$ kpc. This is significantly  shallower than the power spectra
in  earlier measurements including our Galaxy,  the  SMC and DDO 210,
all of which exhibit a slope roughly in the range $-2.5$ to $ -3$
(\citealt{GR93}; \citealt{SSD99}; \citealt{AJS06}).
The shallower slope observed in NGC~628 was interpreted  as arising
from two dimensional (2D) turbulence in the plane of the galaxy's disk
on length-scales  larger than the galaxy's scale-height. The steeper
slope seen  in the other galaxies was ascribed to three dimensional
(3D) 
turbulence at small length-scales.  We expect a transition 
from 3D  to 2D turbulence at a length-scale corresponding to the
galaxy's scale height. An observational detection of this
transition provides a method to  estimate the scale-height of face-on
galaxies \citep{PK01,EK01}, a quantity which is otherwise difficult to
measure. 

NGC~1058 is an almost face-on,  type Sc, spiral galaxy with an
inclination angle in the range  $4^{\circ}$ to $ 11^{\circ}$
\citep{PR07}. In this Letter we use a break in the observed 
HI power spectrum to determine the galaxy's 
scale-height. 
The distance to this galaxy is
uncertain with previous estimates ranging from $10.0 \, {\rm Mpc}$
\citep{B81} to $ 14.5 \, {\rm Mpc}$ \citep{ST74}. Throughout this
paper we adopt a  distance of 10 Mpc  for NGC~1058. At this distance
1$^{\prime\prime}$ corresponds to 48.5 pc.

\begin{figure}
\begin{center}
\mbox{\epsfig{file=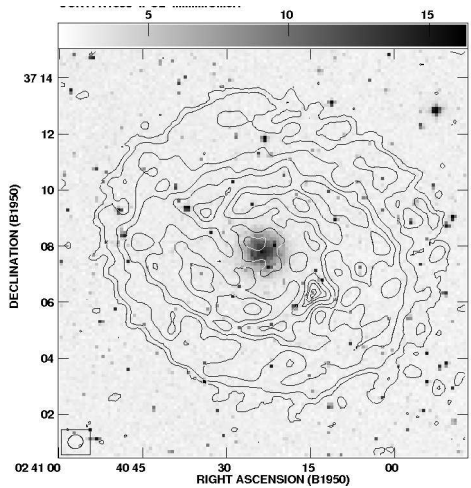,width=2.5in,angle=0}}
\end{center}
\caption{The $13.6^{\prime\prime} \times 13.6^{\prime \prime}$ resolution
  integrated HI column  
density map of NGC~1058 (contours).  The contours levels 
are 3., 10., 30., 50., 70., 100., 120., 130., and 140. $\times
10^{20}$ cm$^{-2}$. 
}
\label{fig:mom0}
\end{figure}

\section{Data and Analysis}
\label{ref:data}

We have used  archival HI data  of  NGC~1058  from the Very Large
Array (VLA). The   observations had been carried out   on 14$^{th}$
June 1993 in the  C configurations of the VLA.
The data was downloaded from the VLA archive and
reduced in the usual way using  standard tasks in classic
AIPS\footnote{NRAO Astrophysical Image Processing System,  
a commonly used software for radio data processing.}. 
The HI emission from NGC~1058 spans 24   channels (i.e, channel
53 to channel 76) of the 127 channel spectral cube. We have used only 
the central $16$ channels (i.e, channel 57 to channel 72) with
relatively strong HI emission. The frequency width of each channel
corresponds to  $2.58~ \, {\rm km~s}^{-1}$.   
 The continuum from the galaxy was subtracted from the data in 
the {\it{uv}} plane using the AIPS task UVSUB.
The resulting continuum subtracted data was used for the subsequent
analysis. Figure \ref{fig:mom0} shows a  total HI column density
(Moment 0) map of NGC~1058 made from this data. 
The HI disk of the galaxy is nearly  face-on. 
The angular extent of the HI distribution in Figure \ref{fig:mom0} 
is measured to be $11' \times 13'$  at a  column density of $10^{19}
\, {\rm atoms \, cm}^{-2}$, which is $\sim 4$ times it's
Holmberg diameter \citep{PR07}.

\begin{figure*}
\begin{center}
\includegraphics[angle=-90]{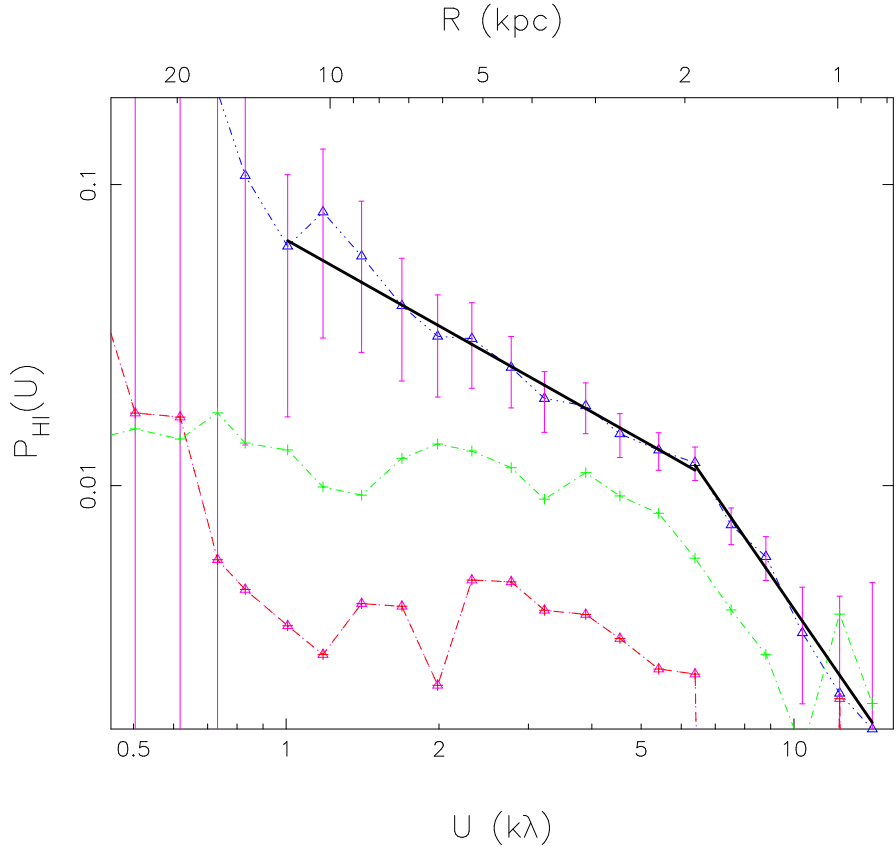}
\caption{ Real (blue, solid)  and imaginary (green, dot-dash) parts of the
 observed value of the HI  
 power spectrum estimator $\hat{P}_{\rm HI}(U)$ for $N=16$.
  $1 \sigma$ error-bars are  shown only   for the real part. The real
 part of the $\hat{P}_{\rm HI}(U)$ from line-free channels (red, bold dot-dash)
  is also shown. }
\end{center}
\label{figure:f2}
\end{figure*}

\citet{AJS06} and \citet{DBBC09} contains a detailed discussion of 
the visibility based HI power spectrum estimator $\hat{\rm P}_{\rm
  HI}(\vU) =
\langle\ \V_{\nu}(\vU)\V^{*}_{\nu}(\vU+\Delta \vU)\ \rangle$, hence we
present only a brief discussion here. 
Here $\vU$ refers to a baseline,  the antenna  separation projected 
in the plane perpendicular to the direction of
  observation,  measured in units of the observing  wavelength
  $\lambda$. Note that, $\vU$ expressed  this way is a dimensionless 
  quantity. It is common practice to express $\vU$ in units of kilo
  wavelength  (k$\lambda$).
Every visibility $\V_{\nu}(\vU)$
 is correlated with all other visibilities 
$\V^{*}_{\nu}(\vU+\Delta \vU)$  within a disk $|\Delta \vU| < (\pi
 \theta_{0})^{-1}$, where $\theta_0$ is the angular extent of the
 galaxy.   
The correlations are averaged over different   $\vU$
directions assuming that  the signal is statistically isotropic in the
plane of the galaxy's image .
To increase the signal to noise ratio we further average the
correlations in  bins of $U$ and over all  frequency channels with  HI
emission.  The expectation value of the estimator $\hat{\rm P}_{\rm
  HI}(\vU)$ is real,  and it is 
the convolution of the HI  power spectrum $P_{\rm HI}(U)$ with  a window 
function  $|\tilde{W}_{\nu}(U)|^{2}$.  The window function 
quantifies the effect of the large-scale HI distribution and the
finite angular extent of  the galaxy, and it can be assumed to be  
sharply peaked around $U=0$ with a width of order
$(\pi\theta_{0})^{-1}$. At baselines $U  
\gg (\pi\theta_{0})^{-1}$ the function  $|\tilde{W}_{\nu}(U)|^{2}$ is 
well 
approximated by a Dirac delta function and the expectation value of
$\hat{\rm P}_{\rm HI}(\vU)$ gives an estimate of the HI power spectrum 
$P_{\HI}(U)$. The effect of $|\tilde{W}_{\nu}(U)|^{2}$ in
estimating the power spectrum and the considerations for determining
$(\pi \theta_{0})^{-1}$ is discussed in detail in \citet{DBBC09}. 

The $1-\sigma$ error-bars for the estimated power spectrum is a sum,
in quadrature,  of 
contributions from two sources of uncertainty. At small $U$ the
uncertainty is dominated by the sample variance which comes from the
fact that we have a finite  and limited  
number of independent estimates of the true power spectrum. At
large $U$, it is dominated by the system noise in each visibility. 

To determine if the slope of the HI power
  spectrum changes with the width of the frequency channel,   we have
  combined $N$ successive channels to obtain a data set with $16/N$
  channels of width $N  2.58 \times \, {\rm km~s}^{-1}$    each. 
We have determined the HI power spectrum for a range of $N$ values.

\section{Results and Discussion}
\label{ref:discuss}

Figure \ref{figure:f2} shows the real and imaginary parts of 
$P_{\rm HI}(U)$,  which is the observed value  of the estimator
$\hat{P}_{\rm HI}(U)$.   As
expected  from the theoretical 
considerations (\citealt{AJS06}), the imaginary part is well
suppressed compared to  the real part. To test 
for  a possible contribution from residual continuum, we also show the
real part of  $P_{\rm HI}(U)$  estimated using line free channels. This is
found to be much smaller than the signal.  For the channels with HI
emission, the observed $P_{\rm HI}(U)$
may be directly interpreted as the HI power spectrum at $U$ values
that are considerably larger than $(\pi \theta_0)^{-1}=0.1 \, {\rm k}
\lambda$.

The power  spectrum is well fitted by
two different power laws $P(U)=AU^{\alpha}$, one  with 
$\alpha\ =-\ 1.0\pm 0.2$ for $U = 1.0
\, {\rm k}  \lambda$ to  $6.5 \, {\rm k} 
\lambda$   (large length-scales) and another with  
$\alpha\ =-\ 2.5\pm 0.6$ for  $U = 6.5
\, {\rm k}   \lambda$ to  $16.0 \, {\rm k}  \lambda$  (small  length
scales). The results are tabulated in Table~\ref{table:t1}.
The best-fit power-law were determined by  a $\chi^2$ minimization
with respect to the amplitude, A and the power-law index, $\alpha$
(\citealt{AJS06}).  The $1-\sigma$ error-bar on $\alpha$,
the parameter of interest here, is determined by projecting 
the ellipse corresponding to $\Delta \chi^2 \ = 1$ onto the $\alpha$
axis in the $\alpha \ -  \ A$ parameter space (page 694 of
\citet{nrcp}).

Both HI density fluctuations as well as spatial fluctuations in the
velocity of the HI gas contribute to fluctuations in the HI specific
intensity.
 Considering a turbulent ISM, \citet{LP00} have shown that it is
 possible to disentangle these two contributions  by studying 
the behavior of the HI power spectrum as the thickness of the
frequency  channel is varied. The  slope of the HI power spectrum  
is  expected to change with the frequency channel thickness if 
it  is due to the gas velocities, However, \citet{AJS06},
\citet{DBBC08}  and \citet{DBBC09} do not find this   
 change for any of the galaxies   in their  sample of spiral and
 dwarf galaxies.   
Further, \citet{EK01}  also reported a similar behavior for  LMC. We 
find that the  HI power spectrum of NGC~1058 does not exhibit a
statistically significant change with increase of the channel
thickness (Table \ref{table:t1}),  
indicating that the observed power spectrum is due to HI  density
fluctuations. 

\begin{table}
\centering
\begin{tabular}{|c|c|c|c|}
\hline 
Baseline Range& N & $\Delta V$   &$\alpha $ \\ 
 (k$\lambda$)      &    & (km s$^{-1}$) &           \\ 
\hline
 \hline 
             & 1  & 2.68  & $-1.0\pm 0.2$ \\
$1.0 - 6.5 $ & 8  & 21.44 & $-0.9\pm 0.2$ \\
             & 16 & 42.88 & $-0.8\pm 0.2$ \\
\hline
            & 1  & 2.68  & $-2.5\pm 0.6$ \\
$6.5 - 16.0$ & 8  & 21.44 & $-2.2\pm 0.6$ \\
            & 16 & 42.88 & $-2.2\pm 0.5$ \\
\hline
\end{tabular}
\caption{NGC~1058 power spectrum are well fitted by  two power laws of
  $P_{HI}=AU^{\alpha}$ in two different baseline ranges.
 This table   summaries the result for different channel width.} 
\label{table:t1}
\end{table}

A baseline $U$  corresponds to an angular scale of $1/U$ in the plane
of the sky. In the subsequent discussion, we use $D=10 \ {\rm Mpc}$
(distance to the   galaxy)  to convert the baseline $U$ to a
length-scale $D/U$ in the plane of the galaxy's image. We find that
the slope   
of the  power   spectrum of NGC~1058 is 
$\alpha\ =-\ 2.5\pm 0.6$   at small length-scales  ($0.6  -  1.5 \,
{\rm kpc}$) whereas  it is 
$\alpha = -1.0 \pm 0.2$  
at large length-scales ( $1.5 - 10.0 \, {\rm    kpc}$).
The largest  length-scale   ($10 \, {\rm kpc}$)  is
definitely larger than the  typical HI scale heights within the
Milky-Way \citep{LH84,WB90} and in  external spiral galaxies
(e.g. \citealt{narayan}). It is then quite reasonable to 
conclude that   the slope  $\alpha = -1.0 \pm 0.2$ is that of 2D
turbulence  in the plane of the galaxy's disk. 
Simulations of the power spectrum of HI emission from a  face-on disk
galaxy \citep{DBBC09} predict  a  change in the slope 
$\alpha$ corresponding to a  transition from 2D to 3D  turbulence.
This is expected   to occur at a baseline $U=D/\pi  z_h$, where $z_h$
is the galaxy's scale-height.  Interpreting the 
change in the  slope  of the observed power spectrum  seen at $U=6.5
\, {\rm k} \lambda$ as the transition from 2D to 3D turbulence, we 
find the scale-heigth of NGC~1058 to be $490 \, {\rm pc}$. 
Further, we interpret the slope of $\alpha\ =-\ 2.5\pm 0.6$ to be 
that of 3D turbulence. To  our knowledge this is
the  first observational determination of the scale-height of a nearly
face on spiral galaxy through its HI power spectrum. 
The width of the bin $\Delta U$, corresponding to the baseline where
the break occurs, gives an estimate of the uncertainty in the
estimated scale-height as $\Delta z_{h} = D \Delta U/\pi U^{2}$, 
whereby we have $490  \pm 90 \ \rm {pc}$. \cite{kregel04} present HI
images of a large sample of edge on intermediate 
to late type spirals; from their data the ratio of the HI disk height to 
the radius of the HI disk (at a column density of 1$M_\odot$pc$^{-2}$) is
$\sim 0.06 \pm 0.015$. Since the disk of NGC~1058 has a diameter of $35
\ \rm{kpc}$ (Figure \ref{fig:mom0}) at this column density, we expect
it's scale-height to be $1.05 \pm 0.26 \ \rm{kpc} $,  in
excess of  our result. Note that the  HI scale-height is known to
increase with the  distance from the center, and our result  is an
average  value over the entire disk with possibly a larger
contribution from the inner parts  where the emission is strongest.

A similar behavior of a change in the slope of the power spectrum at larger
length scales has been also observed in LMC (Elmegreen et al. 2001).
The HI power spectrum of the LMC flattens  
at large length-scales, which was interpreted
as a transition from  
three-dimensional  to two-dimensional turbulence. Using this method  
the  scale-height of LMC was found to be 100 pc (\citealt{EK01}).
 It is well known (eg. \citealt{ES04I}) that the slope of the velocity
 power spectrum   changes by unity from  $-11/3$ to $-8/3$ in going
 from 3D to 2D  for incompressible Kolmogorov turbulence
 \citep{K41}. This may be 
 related to the change in slope from $-2.5$ to $-1.0$ measured in 
NGC~1058, though the turbulence here is compressible and the
 observations are those of density fluctuations and not the velocity. 
The relation between the power spectrum of density fluctuations and
 the  velocity power spectrum is not very clear for compressible
 turbulence. 

There is now mounting evidence that spiral galaxies exhibit 
scale-invariant density fluctuations that extend to length-scales 
of  $\sim 10 \, {\rm kpc}$  (eg. NGC~628, NGC~1058) which is
comparable to the diameter of the HI disk. While a large variety of
possible energy sources like proto-stellar winds, supernovae, shocks, etc. 
have been proposed to drive turbulence  \citep{ES04I},  it is still to
be seen whether these are effective  on length-scales as  large
as $10 \, {\rm kpc}$.

It is interesting to note that a break in the  power spectrum, like
the one seen here for NGC~1058, has also been observed in the 
shell-type supernovae remnant Cassiopeia~A \citep{RBDC08}.
The slope of the power spectrum changes from $-2.22 \pm 0.03$ at  
$1.6 - 10 \, {\rm  k}\lambda$ (large length-scales)  to 
$-3.23 \pm 0.09$ at  $11 - 30 \, {\rm  k} \lambda$ (short
length-scales).  This change in the slope is interpreted as a
transition from 2D to 3D magneto hydrodynamic turbulence. The
transition occurs   at  a length-scale  of $50 \, {\rm  pc}$ which
corresponds  to  the thickness of the shell. 

\section*{Acknowledgments}

P.D. is thank full to Sk. Saiyad Ali, Kanan Datta, Tapomoy Guha Sarkar,
Suman Majumder, Abhik Ghosh and Prakash Sarkar for use full
discussions. P.D. would like to acknowledge SRIC, IIT, Kharagpur for
providing financial support. S.B. would like to acknowledge financial
support from BRNS, DAE through the project 2007/37/11/BRNS/357.
The data presented in this paper were obtained from the National Radio
Astronomy Observatory (NRAO) data archive. The NRAO  
is a facility of the US National Science Foundation operated under
cooperative agreement by Associated Universities, Inc.

\end{document}